\documentclass[11pt]{article}
\pdfoutput=1

\usepackage{amsmath, amsfonts, amssymb}
\usepackage{comment}
\usepackage{graphicx}
\usepackage{psfrag}
\usepackage[usenames,dvipsnames,svgnames,table]{xcolor}
\usepackage{enumerate}
\usepackage{soul}
\usepackage{subfig}
\usepackage{cite}
  \usepackage{a4wide}
  
  \usepackage{color}
  \definecolor{dark-gray}{gray}{0.20}
  \definecolor{gray}{gray}{0.30}
  \definecolor{light-gray}{gray}{0.80}
  \definecolor{dark-red}{rgb}{0.7,0,0}
  \definecolor{dark-green}{rgb}{0.1,0.4,0}
  \definecolor{dark-blue}{rgb}{0.3,0.3,0.7}
  \definecolor{light-blue}{rgb}{0.8,0.8,1}

\usepackage{hyperref}
\usepackage{setspace}
\hypersetup{
	colorlinks=true,
	linkcolor=dark-blue,
	citecolor=dark-red,
	urlcolor=dark-green,
	linktoc=page
}
\newcommand{\be}{\begin{equation}}
\newcommand{\ee}{\end{equation}}

\usepackage{cleveref}
\def\be{\begin{equation}}
\def\ee{\end{equation}}
\def\bea{\begin{eqnarray}}
\def\eea{\end{eqnarray}}

\newcommand{\w}{\wedge}
\newcommand{\vol}{\text{vol}}
\newcommand{\f}[2]{\frac{#1}{#2}}
\renewcommand{\Im}{\text{Im}\,}
\renewcommand{\Re}{\text{Re}\,}

\newcommand{\VVH}[1]{{\color{purple}{\textbf{VVH: }#1}}}

\newcommand{\e}{\textrm{e}}

\newcommand{\dd}{\mathrm{d}}

\def\simleq{\; \raise0.3ex\hbox{$<$\kern-0.75em
      \raise-1.1ex\hbox{$\sim$}}\; }
   \def\simgeq{\; \raise0.3ex\hbox{$>$\kern-0.75em
      \raise-1.1ex\hbox{$\sim$}}\; }

\numberwithin{equation}{section}

\begin{document}

\begin{center}

{\LARGE {\bf  A 10d view on the KKLT AdS vacuum and uplifting}} \\

\vspace{1.5 cm} {\large  F. F. Gautason, V. Van Hemelryck, T. Van Riet and V. Venken }\\
\vspace{0.5 cm}  \vspace{.15 cm} {Institute of Theoretical Physics, KU Leuven,\\
Celestijnenlaan 200D B-3001 Leuven, Belgium
}

\vspace{0.7cm} {\small \upshape\ttfamily  ffg, vincent.vanhemelryck, thomas.vanriet @ kuleuven.be, victoria.venken@gmail.com
 }  \\

\vspace{2cm}

{\bf Abstract}
\end{center}

{\small We analyse the ten-dimensional Einstein equations in the KKLT setting.  We verify that the quartic gaugino term is needed to remove singularities in the on-shell action as suggested by Hamada et.~al. We contrast two approaches that have been taken in the literature when employing the effect of gaugino condensation in the ten-dimensional equations of motion. Here we follow the proposal to insert explicit non-zero fermion bilinar vev into the localised energy-momentum tensor of the 7-branes obtained from varying the 10d off-shell action with respect to the 10d metric.
%emphasize that our approach differs strongly from a similar procedure in which the bilinear vev's are first plugged into the 10d action before taking a variation with respect to the 10d metric. 
Our procedure is common in semi-classical physics and is manifestly local in 10d. However, it does not lead to the KKLT effective field theory. %In particular, our procedure fails to give the exponential dependence of the gaugino condensate on the volume modulus. 
The alternative procedure of deriving the energy momentum tensor after replacing fermion bilinears by the gaugino vev, might be less well motivated in 10d, but reproduces the results of the KKLT effective field theory.% and more a reverse-ingeneering in order to incorporate the 4d EFT of KKLT. We remain however agnostic as to what is the correct approach.}

\setcounter{tocdepth}{2}
\newpage

\section{Introduction}

Moduli stabilisation in string theory has a long history and its understanding is crucial for the study of top-down phenomenology. One of the main breakthroughs came in 2003 with the KKLT model \cite{Kachru:2003aw} and in 2005 with the Large Volume Scenario \cite{Balasubramanian:2005zx}. Both methods to stabilise moduli rely on computing leading-order corrections to ten-dimensional supergravity solutions with orientifold sources and three-form fluxes in type IIB supergravity \cite{Dasgupta:1999ss, Grana:2000jj, Giddings:2001yu}. %In this paper we investigate the KKLT model. 
Despite the long history of this field there has never been any genuine top-down understanding of these mechanisms to stabilise moduli. The arguments always involved a mixture between top-down and bottom-up viewpoints. This is not a problem per se, but it can lead to a false sense of freedom to tune parameters in the bottom-up effective field theory. It is therefore desirable to find at least one concrete top-down description of moduli-stabilisation. 

Recently the field of moduli stabilisation has regained attention due to a growing suspicion that string theory may fail to accommodate any de Sitter (dS) vacuum \cite{Brennan:2017rbf, Danielsson:2018ztv, Obied:2018sgi}. This suspicion is mainly based on three things. Firstly on the lack of explicit dS examples string constructions such as classical type II supergravity  and tree-level heterotic string theory \cite{Green:2011cn,Gautason:2012tb,Kutasov:2015eba,Quigley:2015jia} (see \cite{Danielsson:2018ztv} for a review). Secondly, on non-trivial generalisations of the Dine-Seiberg argument \cite{Ooguri:2018wrx} (see also \cite{Hebecker:2018vxz}).\footnote{See \cite{Garg:2018reu} for early suggestions of a refined form of the dS swampland bounds.} And thirdly, on the old debate concerning the quantum breakdown of dS space (due to backreacting particle creation) \cite{Mottola:1984ar, Polyakov:2007mm, Polyakov:2012uc, Bautista:2015nxc, Anderson:2017hts, Dvali:2018jhn, Danielsson:2018qpa}.\footnote{See \cite{Akrami:2018ylq, Kachru:2018aqn, Cicoli:2018kdo, Kallosh:2019axr} for some critical remarks on these ideas. }  

This revival of interest has provided a healthy spark of ideas to construct new possible dS vacua (or more general accelerating cosmologies) \cite{Kallosh:2018nrk,Banerjee:2018qey, Blaback:2018hdo,Cordova:2018dbb, Heckman:2019dsj, Weissenbacher:2019bfb} or revisit existing ones \cite{Moritz:2017xto, Gautason:2018gln, Moritz:2018sui, Bena:2018fqc}. In this paper we direct our attention to the consistency of the KKLT model; both the construction of the AdS vacuum and its potential uplift to a meta-stable dS vacuum.  In particular the uplift procedure has been heavily criticised on two fronts. First the stability of the anti-branes themselves came under intense scrutiny in \cite{Bena:2009xk, Blaback:2012nf, Gautason:2013zw}. It is our opinion that the arguments pointing towards the instability of anti-branes have been to a large extent addressed and refuted in \cite{Michel:2014lva,Cohen-Maldonado:2015ssa, Cohen-Maldonado:2016cjh, Armas:2018rsy} (see however \cite{Bena:2014jaa, Bena:2016fqp}). 
The second concern, raised in \cite{Moritz:2017xto} (see also \cite{Dasgupta:2014pma}), has to do with the interaction between the open string degrees of freedom of the anti-brane and closed string degrees of freedom of the background. This interaction is only possible in a compact model where the two effects cannot be separated indefinitely. 

The authors of \cite{Moritz:2017xto} suggested that it should be possible to study the interaction using classical ten-dimensional supergravity including the energy-momentum of a gaugino condensate on D7-branes that live in the UV.\footnote{Other somewhat orthogonal worries about the consistency of uplifting were expressed in \cite{Sethi:2017phn, Bena:2018fqc}.} However, there is some confusion regarding the gravitational backreaction of the gaugino condensate in ten-dimensional supergravity. This was discussed in \cite{Moritz:2017xto, Gautason:2018gln} and we refer the reader to those references for further elucidations. The main observation of \cite{Moritz:2017xto} was that the ten-dimensional energy-momentum tensor behaves such that a dS vacuum is not attainable with a single gaugino condensate.  From a four-dimensional  viewpoint this effect can be understood as a flattening of the potential caused by a backreaction of the supersymmetry-breaking ingredients on the volume modulus. 

Later, it was shown in \cite{Gautason:2018gln} that there is a loophole in the argument of \cite{Moritz:2017xto} due to a subtlety in the computation. Unfortunately the issue could not be settled in \cite{Gautason:2018gln} since some of the terms in the energy-momentum tensor were divergent and a regularisation is needed that could influence the fate of the flattening effects. Recently Hamada \emph{et~al~} \cite{Hamada:2018qef} (see also \cite{Kallosh:2019oxv})  proposed how to perform this regularisation by adding a quartic fermion term to the action. In this note we redo the computations in \cite{Moritz:2017xto,Gautason:2018gln} including this new term. We find that indeed the four-dimensional curvature is expressed entirely in terms of regular \emph{background} fluxes and the gaugino condensate, but we fail to find the KKLT effective theory and explain why. The computation can rather straightforwardly be extended to include anti-branes. 
%and we find that the flattening effects suggested by \cite{Moritz:2017xto} persist, within the approximations made.  There should be an effective field theory description of these flattening effects as suggested in \cite{Moritz:2017xto} and discussed further in \cite{Kallosh:2018wme, Moritz:2018ani, Kallosh:2018psh}. But in this paper we will not address this. 

Simultaneously with this work two other works appeared \cite{Carta:2019rhx, Hamada:2019ack} about the same problem and one of these works \cite{Hamada:2019ack} comes to a very different conclusion then us. We discuss in section \ref{Sec:Discussion} on the relation between our work and these two other papers. The difference relates to manner the energy-momentum tensor is computed and the alternative method of \cite{Hamada:2019ack} and \cite{Carta:2019rhx} does reproduce qualitatively the KKLT effective field theory and finds no dangerous flattening effects when anti-branes are added.

\section{The framework}

The KKLT proposal \cite{Kachru:2003aw} to stabilize moduli in a supersymmetric AdS vacuum rests on reasonable assumptions about quantum corrections to the classical flux compactifications of \cite{Dasgupta:1999ss, Giddings:2001yu}. These corrections have been argued for using four-dimensional effective field theories described within ${\cal N}=1$ supergravity. The quantum correction used by KKLT is the leading non-perturbative quantum effect to the superpotential $W$. Perturbative corrections to $W$ are absent because of non-renormalisation theorems whereas it is argued that (non-)perturbative corrections in the K\"ahler potential $K$ can sometimes be self-consistently ignored.\footnote{The Large Volume Scenario \cite{Balasubramanian:2005zx} uses perturbative corrections to the K\"ahler potential to stabilize the volume modulus at exponentially large value in a non-supersymmetric vacuum.} The non-perturbative quantum correction arises as a result of gauginos living on a stack of D7-branes condensing in a confining vacuum of the \emph{four-dimensional} gauge theory. In what follows we assume, like KKLT, that the underlying Calabi-Yau manifold only admits a single K{\"a}hler deformation. It then follows that there is only one (holomorphic) 4-cycle wrapped by the D7-branes.

At first sight the gaugino condensation obscures a dimensional oxidation of the KKLT model.  However, it has been suggested in \cite{Moritz:2017xto} that it should nonetheless be possible  using ten-dimensional supergravity with D7 probe actions included. The way the gaugino condensate couples to gravity and other closed strings could then be understood simply by keeping explicit non-zero gaugino bilinears of the fermionic part of the D7 action and compute its contribution to the ten-dimensional energy-momentum tensor \cite{Baumann:2010sx,Baumann:2007ah,Dymarsky:2010mf,Heidenreich:2010ad,Koerber:2007xk}. 

Like \cite{Moritz:2017xto} we  compute the trace over the four-dimensional part of the (trace-reversed) Einstein equation. We will perform this computation in detail in next section but here we will discuss the form of the resulting expression,
\begin{equation}\label{R4A}
R_4 = \int_{6}\sqrt{g_6}\, \mathcal{E}[F^2, \lambda^2]\,,
\end{equation}
where $R_4$ is the curvature scalar of the four large dimensions, $\int\sqrt{g_6}$ integration over the compact dimensions and  $\mathcal{E}[F^2, \lambda^2]$ represents some function of the background fluxes ($F^2$) and the gaugino bilinear ($\lambda^2$). We find that our approach is consistent with an AdS spacetime but further relations between the fields are required to make a quantitative comparison with the KKLT result.
%with a length scale of the same order of magnitude as the KKLT vacuum, but with a numerical mismatch. This numerical mismatch is not innocent and most likely signals a failure to reproduce the exponential dependence of the gaugino term on the volume in the KKLT model. 
Strictly speaking the method of the trace-reversed Einstein equation allows to compute only the curvature of a vacuum solution which is related to the on-shell potential. The relation between fields required to make a comparison to KKLT is only obtained from the off-shell potential which we do not derive here.% results and going off-shell requires assumptions relying on the 4D EFT that one should probably not take if one is set out to \emph{derive} the EFT. %Nonetheless we will argue that the exponential dependence cannot be reproduced.

Adding anti-branes to this computation is straightforward as explained in \cite{Moritz:2017xto, Gautason:2018gln}. One can show it simply adds the following term to the right hand side of equation \eqref{R4A}:
\begin{equation}\label{R4B}
R_4 = \int_{6}\sqrt{g_6}\,\bigl( \mathcal{E}[F^2, \lambda^2] -2nT\delta_6 \bigr)\,,
\end{equation}
where $T$ is the warped tension of a stack of anti-D3 branes and $n$ the number of anti-D3's. This seems a contradiction since this term contributes negatively compared to the rest, so how can it uplift? In fact \cite{Moritz:2017xto} realised that, in the right approximation scheme, the anti-brane term is completely subdominant compared to the bulk piece $\mathcal{E}$ and should be ignored. 

This naive ``paradox'' arises because (\ref{R4B}) computes an \emph{on-shell} four-dimensional potential  instead of a full off-shell expression\footnote{See \cite{Moritz:2017xto, Gautason:2018gln} for a more elaborate explanation.}.  The way uplifting should work from this viewpoint is rather different.  Uplifting works by a shift of the volume modulus $\rho$ induced by the supersymmetry-breaking, altering the value of $\mathcal{E}$ and potentially flipping its sign. To make that point very explicit let us for simplicity assume that the term $\mathcal{E}$ has two pieces
\begin{equation}\label{epsilonequation}
\mathcal{E}[F^2, \lambda^2; \rho] = \mathcal{E}_+[F^2, \lambda^2; \rho] + \mathcal{E}_-[F^2, \lambda^2;\rho],
\end{equation}
where $\mathcal{E}_-$ is negative definite and $\mathcal{E}_+$  positive definite and we introduced the symbolic dependence on $\rho$. In the supersymmetric vacuum we obviously have $\mathcal{E}_+[\rho] <-\mathcal{E}_-[\rho]$. But when $\rho$ shifts to $\rho+\delta \rho$ due to supersymmetry breaking we expect that  the value of $R_4$ and therefore $\mathcal{E}[\rho+\delta \rho]$ is less negative. A dS is reached by making sure that $\mathcal{E}_+[\rho+\delta \rho]> -\mathcal{E}_-[\rho+\delta \rho]$.  

Apart from the $\rho$-dependence the only quantities appearing in $\mathcal{E}$ are the \emph{background} fluxes and gaugino vev. We can assume that their values have not altered when breaking supersymmetry. This is a crucial assumption that is at the core of the KKLT model. In other words the fluxes $F^2$ are the ones of the GKP solution in the background. One could worry about this assumption since there are two sources for deviations from the GKP solution: 1) near the anti-D3 branes non-ISD fluxes are sourced whereas the original GKP construction only allowed for ISD fluxes \cite{Giddings:2001yu}. But as we emphasised earlier this is a local backreaction effect, and any extra terms in the Einstein equation that are otherwise missed in the approximation will only contribute negatively.  2) Near the 7-branes also non-ISD fluxes are sourced. Even worse, these fluxes are divergent, similar to what was believed to be the case for anti-D3 branes \cite{Bena:2009xk, Gautason:2013zw} but later shown not too happen because of brane polarisation \cite{Cohen-Maldonado:2015ssa, Cohen-Maldonado:2016cjh}. In fact the discussion of anti-D3 flux singularities parallels that of D7 gaugino-induced flux singularities. Whereas brane polarisation cures the singularities for anti-D3 branes, it was shown in \cite{Hamada:2018qef} that quartic gaugino-terms, previously ignored, do the same for D7-branes. 

The essence of this paper is the computation of the right hand side of equation (\ref{R4B}) incorporating the quartic gaugino term of \cite{Hamada:2018qef} and is outlined in the next section.  

\section{A subtlety in the semi-classical Einstein equations}
At this point a note of caution is required. While this paper was finalised we learned of two other papers \cite{Hamada:2019ack, Carta:2019rhx} that appeared about the same problem. Their results differ from ours and the reason is rather straightforward. The difference relies in how one computes the semi-classical Einstein equation in 10-dimensions:
\begin{equation}\label{Einstein}
	G_{\mu\nu} = \langle T_{\mu\nu}\rangle.
\end{equation}
Our method relies upon the following procedure. We first compute the form of the tensor $T_{\mu\nu}$ in 10 dimensions by varying the action with respect to the 10d metric \emph{keeping all other fields fixed} and keep the fermion bilinears arbitrary. Subsequently we use the 4d gauge theory results for the vevs of the fermion bilinears to fix the expression of $T_{\mu\nu}$. Notice that since the 4d vevs of the fermion bilinears have an explicit dependence on the volume of the compact space, there is an implicit dependence on the 10d metric. Hence this procedure will lead to different result than if one instead first replaces the fermion bilinears in the 10d action with the 4d vevs and \emph{then} compute the variation of the action with respect to the volume. 
In Section \ref{comparison},  we  demonstrate that this difference in the order of variations leads to the results of Hamada et al \cite{Hamada:2019ack}. Note that our method gives a manifest local description in 10d whereas there seems to be no ten-dimensional, manifestly local, covariant description in the approach of \cite{Hamada:2019ack, Carta:2019rhx}. 

The main advantage of the approach followed by \cite{Hamada:2019ack} is that it seems to directly reproduces the 4d computation done by KKLT. In this work we prefer to remain agnostic about the correctness of the 4d result for multiple reasons which we now explain. First of all the motivation of our paper is to establish (or refute) the KKLT model from a 10d perspective, before and after uplift. Hence we cannot assume the 4d EFT as a prior. Alternatively one can regard the spirit of this work to be a test for the proposal that a 10d derivation of KKLT can be found using fermion bilinear vevs in 10D as the only relevant extra contribution to the ``GKP" background. 
Whether semi-classical 10d supergravity with localised 7-brane sources is the appropriate description of the four-dimensional gauge theory in its confining phase is an open question. In this paper we report our results assuming that for the sake of determining the vacuum, it is adequate. Secondly, the precise way in which gaugino condensates couple to gravity is not understood, even in four dimensions. KKLT makes a proposal to simply sum the background GVW superpotential with the field theory superpotential. This is reasonable, but we are not aware of a derivation of this and caveats can be around the corner. For instance, since gaugino condensation is a consequence of long-wavelength fluctuations of fields, one expects it to be sensitive to the background it is supposed to live on. Hence gaugino condensation can look different in AdS or some other curved space. In other words: how exactly gaugino condensation couples to gravity can be more complicated than simply adding it into the supergravity F-term. Furthermore, the classical gravity background breaks supersymmetry via its fluxes. This means that there is an explicit SUSY breaking occuring in theory once we incorporate the gravitational degrees of freedom and hence the assumptions of having SUSY of the off-shell Lagrangian, required for computing gaugino condensation, is not obvious. In what follows we lay out the details of the computation of the 4d piece in the 10d Einstein equation (\ref{Einstein}). As we will show, our results are not incompatible with an AdS solution  but we are unable to determine a quantitative value for the  vacuum energy that can be compared with the KKLT results. This will be  discussed further in  Section \ref{Sec:Discussion}.

\section{The four-dimensional cosmological constant}\label{4dCC}
In this section we utilise the approach of \cite{Giddings:2001yu} to compute the four-dimensional curvature when compactifying ten-dimensional type IIB supergravity with fluxes and D7-brane sources with non-vanishing gaugino bilinear. In doing so, we assume a maximally symmetric four-dimensional spacetime. The ten-dimensional metric reads \footnote{\label{foot:4DMetric} Note that the four-dimensional metric $\dd s_4^2$ is not in four-dimensional Einstein frame because the volume modulus has not been properly factored out. This does not cause any issue with our computation in this section since we assume that all moduli are stabilised by the gaugino condensate. We will return to this issue when we make an explicit comparison to KKLT in section \ref{sec:MatchingTo4DEFT}.}
\be\label{metric}
\dd s_{10}^2 = \e^{2A}\dd s_4^2 + \e^{-2A} \dd s_6^2~,
\ee
where $\dd s_4^2$ is a maximally symmetric on a four-dimensional spacetime, with a constant curvature scalar $R_4$. The warp factor $A$ only depends on the coordinates of the transverse manifold $M_6$ equipped with the metric $\e^{-2A}\dd s_6^2$. Maximal symmetry demands that the ten-dimensional axion-dilaton $\tau$ and 3-form field strength $G_3$ only have legs and dependence along $M_6$. The type IIB 5-form $F_5$ is self-dual and given by
\be \label{Ansatz2}
F_5 = (1+\star_{10})\dd C_4~,\quad C_4 = \alpha\, \vol_4~,
\ee
where $\alpha$ is a function on $M_6$. The inclusion of three-form fluxes induces three-brane charge that contributes to the tadpole which must be cancelled by the inclusion of local sources such as O3/O7 planes. Taking these ingredients into account, \cite{Giddings:2001yu} found a consistency condition demanded by the trace-reversed Einstein equation that takes the form\footnote{Here $\triangle_6 \Phi^- = -\star_6\dd\star_6\dd \Phi^-$ and all inner products of forms are ten-dimensional, i.e. they include warp factors.} 
\be \label{GKP}
\triangle_6 \Phi^- =  R_4 + \f{\e^{2A}}{\Im \tau}|G_3^-|^2 + \e^{-6A}|\dd \Phi^-|^2~,
\ee
where
\be
\Phi^\pm = \e^{4A} \pm \alpha~,\quad G_3^\pm = \f12 (\star_6 \pm i)G_3~.
\ee
This equation can also be derived directly from the type IIB action (together with O3/O7 source terms) upon varying with respect to the warp factor $A$ (see appendix \ref{Appendix_warp_variation}). The power of this equation becomes apparent when one integrates \eqref{GKP} over $M_6$. The left-hand-side integrates to zero which implies that the four-dimensional curvature scalar $R_4$ is non-positive. The GKP vacua are the ones with $R_4=G_3^-=\Phi^-=0$. 

We now derive the equation corresponding to \eqref{GKP} when $N>1$ D7-branes are included with non-vanishing fermion bilinear. At this stage we will be agnostic about the value of the fermion-bilinear, and assume that it will take some non-vanishing value at low energies where the gauge group condenses. In the procedure outlined we combine the ten-dimensional type IIB action with the effective action of  D7-branes. The bosonic D7-brane action
\be\label{D7DBI}
S_{\text{D7}} = -2\pi N\int_{\Sigma_8} \dd^{8}x (\text{Im}\tau)^{-1}  \sqrt{|P[g]|}+2\pi N\int_{\Sigma_8} P[C_8]~,
\ee
does not contribute to \eqref{GKP} since ``it is BPS'' with respect to the background.\footnote{BPS is used in a loose form since fluxes can already break supersymmetry of the GKP background.} One can verify this claim by using the ansatz \eqref{metric} and noticing that the warp factor drops out. Even though the bosonic action does not play an important role in our discussion, the fermion terms do. Indeed the gauge theory living on the D7-brane world volume descends to an ${\cal N}=1$ gauge theory in four dimensions with various matter couplings. In particular the eight-dimensional world-volume fermions give rise to the ${\cal N}=1$ gaugini in four dimensions.\footnote{We refer to \cite{Moritz:2017xto,Gautason:2018gln} for an explicit map beween the eight-dimensional fermions and the gaugino.} The fermionic D7-brane action contains an interaction term beween the gaugino bilinear $\lambda\lambda\equiv \text{tr}( \lambda^{\dot\alpha}\lambda_{\dot \alpha})$ and the three-form  $G_3$ \cite{Martucci:2005rb} (see \cite{Camara:2004jj} for an early discussion)
\be\label{definition_I}
S_\text{D7}^\text{ferm} =\pi \int \star_{10}\left(G_3\cdot I + \overline{G_3}\cdot \bar{I}\right)~,\qquad I = \frac{e^{-4A}}{\sqrt{\Im \tau}} \frac{\bar{\lambda}\bar{\lambda}}{16\pi^2}  \Omega~\delta(\Sigma_8)\,.
\ee

The contribution of the fermionic terms to \eqref{GKP} turns out to be singular due to the backreaction of the fermion bilinears on the three-form fluxes\cite{Moritz:2017xto,Gautason:2018gln}
\be
G_3 \sim (\Im \tau) \bar{I}~,
\ee
where $I$ is the source appearing in \eqref{definition_I} and carries an explicit delta function. In fact, even the on-shell action is UV divergent due to this singular backreaction. In \cite{Hamada:2018qef} it was suggested that the singular backreaction is ultimately a result of omitting  a four-fermion term in the D7-brane action. Currently the four-fermion terms are not known for the D-brane action and so \cite{Hamada:2018qef} suggested to fix their form such that the singularities would be cancelled. The suggestion of \cite{Hamada:2018qef} was inspired by a similar problem appearing in Ho{\v r}ava-Witten theory\cite{Horava:1996ma,Horava:1995qa,Horava:1996vs} which was resolved by a four-fermion term that appears together with the M-theory 4-form in a perfect square. This was in direct analogy with a similar square structure that appears in heterotic supergravity \cite{Dine:1985rz, Kallosh:2019oxv}. It is therefore reasonable to assume that the fermionic D7-brane coupling \eqref{definition_I} together with the kinetic terms for $G_3$ should be combined with the four-fermion terms to form a perfect square \cite{Hamada:2018qef,Kallosh:2019oxv}. Indeed this goes a long way to prevent singular on-shell action as a result of backreaction the fermion bilinear. 

The perfect square replaces the original action for $G_3$ and the fermionic coupling by
\begin{align}
\label{replacementG3toPerfectSquare}
S_3 = -\pi \int\star_{10} \f{\left|G_3 - (\Im \tau)\mathcal{P}(\bar{I})\right|^2}{(\Im\tau)}  \,,
\end{align} 
The projector $\mathcal{P}$ is defined to eliminate the coexact piece of the form it acts on. It does not affect the $G_3$ equation of motion but is the final piece of the puzzle to ensure that the on-shell action is regular \cite{Hamada:2018qef}. %We discuss the projector in more detail in \cref{projectorsec} \FFG{Do we need it?}. 
Upon expanding the square the action consists of three terms:
\be
S_{3}
= -\pi \int \star_{10}\f{\left|G_3\right|^2}{(\Im\tau)} + \pi\int \star_{10} \left(G_3 \cdot \mathcal{P}(I) + \overline{G_3}\cdot \mathcal{P}(\bar{I}) \right)- \pi\int \star_{10} (\Im \tau) |\mathcal{P}(I)|^2\,.
\ee
The first term is the standard bulk term for $G_3$, the second term is the fermionic action of the D7-brane considered before in \cite{Gautason:2018gln,Moritz:2017xto,Dymarsky:2010mf}, with the exception that the three-form $I$ is now projected to the set of closed forms. The last term is the new quartic fermion term. We are now ready to re-derive the equation \eqref{GKP} with the D7-brane fermion contribution. The explicit computation is carried out in \cref{Appendix_warp_variation} and the result is
\be
\label{Corrected_GKP_equation}
\triangle_6 \Phi^- ={R}_4 + e^{-6A}|\dd \Phi^-|^2  + e^{2A}\left(\f{|G_3^-|^2}{ \Im \tau} - \f{3}{4}\Re \big(G_3\cdot {\cal P}(I)\big) + \f{\Im\tau}{4} |{\cal P}(I)|^2\right)~.
\ee
This expression is exact and despite its appearance is completely regular.\footnote{Equation \eqref{Corrected_GKP_equation} does have delta-function sources on the right-hand-side but is regular upon integration. This should be compared to the situation without quartic fermion terms where even the integrated expression was singular due to $\delta^2$ terms appearing.} In order to demonstrate this and further analyse equation \eqref{Corrected_GKP_equation} we must now make simplifying approximations. 

First we rewrite the delta function appearing in the source term $I$ in terms of a Green's function. Schematically we write
\be
\f{\partial^2G}{\partial z\partial\bar z} = \delta(\Sigma_8) - \frac{1}{\mathcal{V}_2}~,
\ee
where $z$ and $\bar{z}$ are complex coordinates on the transverse two-cyle and $\mathcal{V}_2$, a volume factor of that two-cycle, is introduced to make this equation consistent. %One can see that it is needed by integrating both sides of the equality; since the left-hand-side is a total derivative, it gives zero and so the right-hand-side must also give zero. 
We see that the 3-form $I\sim\delta(\Sigma_8)\Omega$ is not closed and so the 
role of the projector $\mathcal{P}$ in this case is to add terms of the form $(\partial_z\partial_z G) \Omega$ such that the resulting expression is closed. Explicitly we have\cite{Hamada:2018qef}
\be\label{PIdef}
\mathcal{P}( \bar{I}) = \frac{e^{-4A}}{\sqrt{\Im  \tau}} \frac{{\lambda}{\lambda}}{16\pi^2}  \left(\dd \f{\partial G}{\partial z} \wedge \bar{\Omega}_2 + \frac{1}{\mathcal{V}_2} \bar{\Omega}\right)\,,
\ee
where $\Omega_2$ is the holomorphic two-form on the holomorphic four-cycle. Notice that we have made a crucial assumption here which we will continue making in the following. Namely we have assumed that close to the seven-branes where most of our analysis takes place, we can treat the warp factor as constant. Furthermore we assume the so-called Sen limit \cite{Sen:1996vd} where the stack of 7-branes consists of an O7 with 4 parallel D7-branes on top, in this case the 7-brane stack does not source a gradient for $\tau$ and so it can be taken to be a constant.\footnote{One could wonder whether in this case the orientifold projection does not eliminate the backreaction of the gaugino bilinears on the three-forms. We do not address this question here, but argue that our setup is sufficiently close to the Sen limit such that the gradients of $\tau$ do not enter without eliminating the three-form backreaction altogether.} These assumptions allow us to determine the backreacted three-forms in the same spirit as \cite{Dymarsky:2010mf}. For notational simplicity we write
\be
\label{defIsingIreg}
\mathcal{P}(I) = I^{\text{sing}} + I^{\text{reg}}\,,
\ee
where $I^{\text{sing}}$ is the exact part of ${\cal P}(I)$ in \eqref{PIdef} which is also singular, and $I^{\text{reg}}$ is the harmonic part which is regular. 
%The combined three-form flux and fermionic D7-brane action takes then the form:
%\begin{align}
%S_{3} &= -2\pi \int \dd^{10}x \sqrt{|g_{10}|}\; \frac{1}{2 \Im \tau} \left|G_3 - (\Im \tau)\left(\overline{I^{\text{sing}}} + \overline{I^{\text{reg}}} \right)\right|^2
%\end{align}
%
%If we define the quantity $X$ to be 
%\be
%X =  i(\text{Im}\tau) e^{4A}\left(\overline{I^{\text{sing}}}+\overline{I^{\text{reg}}}\right),
%\ee
%the $G_3$ equation of motion takes the nice form \cite{Moritz:2017xto}
%\be
%\label{eq:Lambda_EoM}
%\dd \Lambda - \frac{\dd \tau}{\tau-\bar{\tau}} \wedge (\Lambda + \bar{\Lambda}) = \dd X - \frac{\dd \tau}{\tau-\bar{\tau}} \wedge (X + \bar{X}),
%\ee
%where 
%\be
%\Lambda = e^{4A}\star_6 G_3 - i \alpha G_3 = \Phi^+G_3^- + \Phi^- G_3^+\,.
%\ee 
Finally in order to solve for $G_3$ we assume that the background close to the seven-branes is only slightly perturbed from the standard GKP background for which $\Phi^-=e^{4A}-\alpha =0$. By our previous assumption that $\e^{4A}$ is constant, this implies that the Chern-Simons terms in the action does not enter in the equations of motion for $G_3$.
%We now make some approximations. We take $C_4$ (and thus constant $\alpha$) and the warp factor to be constant, and moreover that $\Phi^-=e^{4A}-\alpha =0$ as in a GKP background. Furthermore, we will work in the Sen limit  such that the axion-dilaton is constant. 
With these approximations, the solution to the equation of motion for $G_3$ derived from action \cref{replacementG3toPerfectSquare} takes a simple form\footnote{Under these assumptions, the Chern-Simons action does not contribute to the $G_3$ equations of motion.  The details of the fate of the Chern-Simons action in our set-up are discussed in \cref{CSappendix}. Note that since $I^{\text{reg}}$ is harmonic with these assumptions, it can always be combined with an arbitrary harmonic contribution that can be added to a solution to the equations of motion to produce another solution to the equations of motion and we denote the entire harmonic contribution as $G_3^0$.}
\be
\label{G3solved_Ising_G30}
G_3 = (\Im \tau)\overline{I^{\text{sing}}} + G_3^0\,,
\ee
where $G_3^0$ is harmonic and quantised. We can now evaluate the expression \eqref{Corrected_GKP_equation} which constrains the four-dimensional cosmological constant. Substituting the solution of the $G_3$ equation of motion \eqref{G3solved_Ising_G30}, and after some algebra outlined in \cref{Appendix_warp_variation} we find 
\begin{align}
\label{GKP_regularised}
0=\int\star_6\left( {R}_4 + \frac{e^{2A}}{ \Im \tau}|G_3^{0,-}|^2 -\frac{3e^{2A}}{4}\Re\left(G_3^0  \cdot I^{\text{reg}}\right) + \frac{e^{2A}}{4}(\Im \tau) |I^{\text{reg}}|^2\right)~,
\end{align}
The equation only contains the background flux $G_3^0$ and the fermion condensate through the regular part $I^{\text{reg}}$. We notice that all singular terms have been eliminated from the expression as a result of our approximations. We expect that more generally the expression \eqref{Corrected_GKP_equation} will be regular also when the approximations are relaxed.

Equation \eqref{GKP_regularised} is the main result of our paper. From it we can extract the four-dimensional curvature $R_4$. However we notice that as it stands the curvature is not restricted to be negative for all possible configurations of the fields. In the next section we discuss a schematic comparison to the four-dimensional KKLT vacuum before any anti-D3 branes are included.

\section{Comparing to KKLT}
\label{sec:MatchingTo4DEFT}
We now compare our results with the four-dimensional effective theory considered by KKLT \cite{Kachru:2003aw}. This  theory is specified in terms of a K{\"a}hler potential $K$ and superpotential $W$, given by
\be\label{KKLT}
K = -3\log(2\Im\rho) - \log(2\Im\tau)~,\qquad W = W_0 + {\cal A}~\text{exp}(2\pi i \rho/N)~,
\ee
where $\rho$ is the complex volume modulus, related to the volume of the internal manifold by $\Im \rho \sim  {\cal V}_6^{2/3}$ and $W_0$ is the constant Gukov-Vafa-Witten superpotential (after integrating out the complex-structure moduli). In order to compare our ten-dimensional result \eqref{GKP_regularised} with the effective theory \eqref{KKLT} we translate two results from the four-dimensional effective field theory back to ten dimensions. In particular we view the fluxes $G_3^{0}$ as the ISD ones, before the inclusion of the non-perturbative effects. Therefore $G_3^{0,-}$ vanishes and $G_3^0\cdot \Omega$ can be related to $W_0$\footnote{We would like to thank Pablo Soler for correcting a phase factor convention.}
\be
\label{W0}
\int {\star}_6 (G_3^0 \cdot \Omega) = W_0~.
\ee
%Secondly in the supersymmetric vacuum the four-dimensional F-term equation $D_\rho W = \partial_\rho W  + W\partial_\rho K=0$ implies that $\Re (G_3^0  \cdot I^{\text{reg}})=0$. 
In KKLT the gaugino condensate $\lambda\lambda$ is related to the non-perturbative effects appearing in the four-dimensional superpotential \eqref{KKLT}. We use the standard gauge theory expression for the fermion bilinear in the confining vacuum (here in string units) \cite{Kaplunovsky:1994fg}
\be \label{vev}
\langle \lambda \lambda \rangle=16\pi i ~ e^{K/2} ~\partial_\rho W = -\f{32 \pi^2}{N} \frac{{\cal A}\e^{2\pi i\rho/N}}{(2 ~\Im \rho)^{\frac{3}{2}}} (\Im \tau )^{-\frac{1}{2}}~,
\ee
where, like KKLT \cite{Kachru:2003aw} we used that the complex volume modulus $\rho$ equals the complex gauge coupling of the four-dimensional gauge theory. We also rescaled with the string coupling as in \cite{Hamada:2019ack}. % similarly here $W$ should be the effective superpotential in the IR field theory which (appart from the flux term $W_0$) is identified to the supergravity superpotential \eqref{KKLT}. 
We note that ${\cal A}$ is determined by the dynamically generated scale of the IR theory and is left unspecified. % but it should be determined by the details of the UV theory as well as the choice of vacuum.

We emphasise here that the exponential dependence of the volume modulus is crucial in the 4d EFT and its off-shell scalar potential. When balanced with the small flux contribution $W_0$, this exponential stabilises the K\"ahler modulus at a somewhat large volume and exponentially small values of the 4d scalar potential. The 10d supergravity analysis performed here, where we opted for a local description of the gaugino condensate, does not reproduce this exponential behaviour. This is reminiscent from the fact that this procedure requires to plug in the vev after varying the action with the metric (which only renders polynomial dependences). Nevertheless we proceed in further rewriting eq. \eqref{GKP_regularised}.\footnote{
It is important to note the $\lambda$ appearing in equation \eqref{vev} is a canonically normalised gaugino in Einstein frame. As mentioned in footnote \ref{foot:4DMetric}, the 4d metric must be rescaled in order to go to proper 4d Einstein frame. This amounts to a rescaling of the metric $g_4 \to (\Im \rho)^{-3/2} g_4$ which leads to the rescaling of the curvature $R_4 \to (\Im \rho)^{3/2} R_4$. In this new frame the fermion gaugino term is not canonical and so we must also rescale the gaugino $\lambda \to (\Im\rho)^{9/8} \lambda$ such that $\lambda\lambda \to  (\Im\rho)^{9/4} \lambda\lambda$.
It also amounts to rescaling $\Omega \rightarrow (\Im \rho)^{\frac{3}{4}}\Omega$ and $A\cdot B\rightarrow (\Im \rho)^{-\frac{3}{2}}A\cdot B$ where $A,\;B$ are three-forms.}
Using the map between the four-dimensional theory and the ten-dimensional quantities we find the following result
\begin{align}
\int\star_6\Re \big(G_3^0  \cdot I^{\text{reg}}\big) &= \frac{e^{-4A}\Re\left( W_0  \bar{\lambda}\bar{\lambda} (\Im \rho)^{\frac{3}{2}} \right) }{16\pi^2{\cal V}_2\sqrt{\Im \tau}},%\nonumber \\ &= \frac{- e^{-4A}}{3N{\cal V}_2\sqrt{2~\Im \tau}} \Re\left( i(\Im \rho)^{\frac{3}{4}} |{\cal A}\e^{2\pi i\rho/N}|^2(3+\f{4\pi}{N}\Im\rho) \right)=0~,
\end{align} 
where we made use of eq. \eqref{W0}.

%Using these results our ten-dimensional equation \eqref{GKP_regularised} reduces to
%\begin{align}
%\int\star_6\left({R}_4 + \frac{e^{2A}}{4}(\Im \tau) |I^{\text{reg}}|^2\right)=0 \,.
%\end{align}
More explicitly, we can insert the expression for $I^{\text{reg}}$ in \eqref{PIdef} and make use of the rescalings of $R_4$ and $\lambda\lambda$ as above to find 
\be
\label{eq:R4computed}
{R}_4 =  -\frac{2}{\mathcal{V}_2^2}(\Im \rho)^3\left|\frac{\lambda\lambda}{16\pi^2} \right|^2 + \frac{3}{4}\frac{1}{{\cal V}_2\sqrt{\Im \tau}}\Re\left( \overline{W_0}  \frac{\lambda\lambda}{16\pi^2} \right) . 
%= -\frac{1}{(\Im\rho)N^2}\left|{\cal A}\right|^2\e^{-4\pi \Im\rho/N}\,.
\ee
We will also extract the volume dependence out of $\mathcal{V}_2 = \tilde{\mathcal{V}}_2 (\Im \rho)^{\frac{1}{2}}$ so that we can rewrite eq. \eqref{eq:R4computed} with the explicit volume dependence 
\be
\label{eq:R4computed2}
{R}_4 =  -\frac{2}{\tilde{\mathcal{V}}_2^2}(\Im \rho)^2\left|\frac{\lambda\lambda}{16\pi^2} \right|^2 + \frac{3}{4}\frac{1}{\tilde{\mathcal{V}}_2\sqrt{(\Im \rho)(\Im \tau)}}\Re\left( \overline{W_0}  \frac{\lambda\lambda}{16\pi^2} \right)\,. 
%= -\frac{1}{(\Im\rho)N^2}\left|{\cal A}\right|^2\e^{-4\pi \Im\rho/N}\,.
\ee
Without further information we are unable to relate the two terms in \eqref{eq:R4computed2}. Based on the four-dimensional effective theory, we expect the two terms to be related to each other on shell but without solving the equations of motion we are unable to verify this.
%Defining a function $f^{-1}(\rho) \equiv \overline{W_0}^{-1}(\Im \rho )^{\frac{5}{2}}(\Im \tau)^{\frac{1}{2}}\frac{\bar{\lambda} \bar{\lambda}}{16\pi^2}$, which would be independent of $\rho$ in KKLT at leading order, we can rewrite this equation in a more compact form:
%\be
%{R}_4 =  -(\Im \rho)^2 \left|\frac{\lambda\lambda}{16\pi^2} \right|^2 \left( \frac{2}{\tilde{\mathcal{V}}_2^2} - \frac{3}{4} \frac{1}{\tilde{\mathcal{V}}_2} \Re\left( f(\rho)\right)\right). 
%\ee
In particular, this equation does not directly reveal the sign of the curvature. A negatively curved AdS solution is however certainly not ruled out. % Although it is tempting to use the supersymmetry condition of the 4d KKLT EFT (relating the gaugino condensate vev to the flux superpotential), our 10d analysis does not allow for such a useful relation to simplify this equation further.
One way to progress would be to perform the supersymmetry analysis directly in ten dimensions but this is beyond the scope of our work.
%This final expression is everywhere negative definite. Any terms that could have given a positive contribution have disappeared. Of course, a supersymmetric vacuum should be either AdS or Minkowski, so ${R}_4$ should indeed be negative or zero. Note that we reproduce almost exactly the four-dimensional potential evaluated in the vacuum,
For comparison we give here the EFT value of the curvature obtained from the four-dimensional KKLT model before anti-D3 branes are included (in Planck units):
\be
 R_4^\text{KKLT} = 4V_\text{KKLT} = -12 \e^K|W|^2 = -\frac{16\pi^2}{3} (\Im \rho)^2 \left| \frac{\lambda\lambda}{16\pi^2}\right|^2\,. %-\f{2\pi^2}{3(\Im\rho)N^2}|{\cal A}|^2\e^{-4\pi\Im\rho/N}~.
\ee
%we see that the local 10d treatment of the gaugino condensate does not agree with KKLT. \FFG{I am really confused here because we dont finish the comparison. Is there a way to compare or not?}
%The reason for the numerical mismatch could be an artefact of the approximations made.\footnote{Another possible source for the mismatch is a subtle change of conventions between the compactified gauge theory and the four-dimensional literature.} 
%There however the curvature is calculated by immediately reducing the 10d action to the 4d effective one, plugging in the gaugino condensate vev. It is however our opinion that this approach is not fully honest 10d and given the exponential volume dependences is bound to reproduce KKLT at leading order. We kept a more agnostic view by doing an honest 10d calculation. It is however clear from our analysis that a \textit{local} 10d treatment with the proposal from \cite{Hamada:2018qef} does not result in the KKLT EFT. This begs the question how the KKLT vacuum can be reproduced from 10d non-local analysis. We leave this question open to further research.

%\VVH{I do not think this below is relevant:}\\
%An important approximation we made is that we have put the regular harmonic piece of the 3-form flux equal to the classical GKP solution. This approximation should be very good nearly everywhere in the manifold, but most likely not close to the 7-branes. Hence one could speculate that possible extra term then arise that make the match to the AdS vacuum energy computed in KKLT exact. We leave this for future research. 

Finally, let us recall that the anti-D3 brane tension contributes only negatively to ${R}_4$. An uplifting to dS is supposed to occur through the mediation of gaugino terms that contribute positively to ${R}_4$. Such positive terms can only arise from the term $\Re\big(G_3^0  \cdot I^{\text{reg}}\big)$ being non-zero in a supersymmetry breaking vacuum.

\section{Discussion} \label{Sec:Discussion}

\subsection{Summary and discussion of results}
In this paper, we have revised the computations of \cite{Gautason:2018gln} taking into account the flux renormalisation mechanism provided by the quartic gaugino term on D7-branes, as proposed in \cite{Hebecker:2018vxz}. We must rely on a number of assumptions necessary to directly use the techniques of \cite{Hebecker:2018vxz}, namely constant axion-dilaton, warping, and $C_4$-potential, and most importantly we used that the harmonic 3-form flux is the GKP flux. Consistent with the result of \cite{Gautason:2018gln} we are unable to establish the sign of the four-dimensional curvature from ten-dimensional analysis alone. We explicitly showed that within our assumptions the quartic gaugino terms proposed by \cite{Hamada:2018qef} are sufficient to regularize the singularities encountered in \cite{Moritz:2017xto,Gautason:2018gln}. %Together with four-dimensional supersymmetry constraints we have shown that there are no positive terms  in $R_4$ consistent with the statements in \cite{Moritz:2017xto, Gautason:2018gln}. In the notation of equation (\ref{epsilonequation}) we found that $\mathcal{E}_+=0$. It is quite striking that a no-dS result comes out that manifestly in our approach. %Because of this, it is not possible for supersymmetry-breaking by introduction of anti-D3 branes to provide an uplift to de Sitter space. 
%Of all approximations made the crucial one leading to this result is the approximation made about the harmonic three-form flux. It should be mentioned that our result does rely  on the comparison to the four-dimensional EFT which allowed us to eliminate the mixed term in equation \eqref{GKP_regularised}.

A worthwile direction for future research would be to see if one can still obtain a definite result after when it comes to de Sitter solution using anti-D3 brane uplift terms.  It is tempting to speculate what happens. For that we go back to equation (\ref{Corrected_GKP_equation}). Note that this equation was derived without any of the simplifying approximations and is hence valid in general circumstances. We have argued that the renormalisation due to the four-fermion term is such that the only 3-form fluxes appearing in  (\ref{Corrected_GKP_equation}) are the \emph{background} fluxes, ie those of the classical GKP solution. This means we can put $|G_3^{0,-}|^2 =0$ since the flux appearing will be ISD. This is expected to be broken by the gaugino condensate but the deviations from the ISD background are mostly ``self-energy" that has been removed by the renormalisation procedure. Once anti-branes are added they will also create some local non-ISD fluxes down the throat. But we will neglect these since they anyway come with a negative contribution to $R_4$ and so will not help in getting de Sitter.  The main question then is whether the only positive contribution to the vacuum energy %Secondly it is tempting to suggest that 
\be \label{conjecture}
\int \star_6\,\text{Re}\big(G_3\cdot {\cal P}(I)\big)\,, 
\ee
can dominate over the rest. This question seems to be very difficult to answer, and most likely requires a full solution the entire system of the ten-dimensional equations of motion which surely is out of reach.

%This is a non-trivial statement and, within our assumptions turned out to be a consequence of the supersymmetry of the AdS solution. 
One way to approach this problem is make a precise analysis of the term (\ref{conjecture}) for a supersymmetric configuration. This could be done by analysing the ten-dimensional supersymmetry variations in presence of the gaugino bilinear. %The way one could check whether this is true in general, without making assumptions and without appealing to the four-dimensional supergravity is by analysing the ten-dimensional supersymmetry variations in presence of the gaugino bilinear. 
This is a very interesting exercise which we plan to carry out in the future. %If it proves (\ref{conjecture}) correct then the general result for the four-dimensional curvature is simply
%\be
%R_4 = -\f{1}{\cal V} \int \star_6\frac{e^{2A}}{ \Im \tau}|G_3^{0,-}|^2 -\f{1}{\cal V} \int \star_6\, e^{2A}\frac{\text{Im}\tau}{4}|\mathcal{P}(I)|^2\,,
%\ee
%which is manifestly negative. Anti-branes only add extra negative terms. 

%If this turns out to happen then the only angle to achieve dS is if some of the background fluxes and gaugino phases change when supersymmetry is broken. This goes against the very assumptions about the validity of the four-dimensional effective field theory, which assumes that the complex structure moduli (related to the fluxes) and the axion (related to the gaugino phase) do not shift positions. 

Finally we wish to iterate the statements in \cite{Moritz:2017xto, Gautason:2018gln}: to have certainty that uplifting AdS vacua with small supersymmetry-breaking ingredients does not lead to runaways instead of meta-stable dS vacua, one should achieve the following parametric scaling:
\be\label{swamp2}
m^2L^2 \gg 1\,.
\ee
In this equation $m^2$ is the mass squared of the lightest modulus in the AdS vacuum and $L$ the AdS length. Interestingly not a single clear top-down AdS vacuum in string theory achieves this. The only cases known to us are ``stringy-inspired" like racetrack fine-tuning \cite{Kallosh:2004yh} or non-geometric fluxes \cite{deCarlos:2009fq, Palti:2007pm}. It was conjectured in \cite{Moritz:2018sui} that racetrack fine-tuning is in the Swampland\footnote{See \cite{Blanco-Pillado:2018xyn, Kallosh:2019axr} for some criticism on this.} whereas reference \cite{Gautason:2018gln} conjectured this to be true for all vacua obeying (\ref{swamp2}). Since the conjecture of \cite{Gautason:2018gln} forbidding (\ref{swamp2}) is a statement about AdS vacua and not dS vacua we believe it should be easier to verify than the actual no-dS conjecture of \cite{Brennan:2017rbf, Danielsson:2018ztv, Obied:2018sgi}, and it would count as non-trivial evidence in support of the no-dS conjecture.

\subsection{Comparison with recent papers}\label{comparison}
Simultaneous with the appearance of this work, two other papers appeared discussing the same problem \cite{Carta:2019rhx, Hamada:2019ack}. Reference \cite{Hamada:2019ack} set out to do similar computations as done here but came to a different conclusion. %The reason can be traced to a different approach in handling the gaugino condensate in ten dimensions. 
We differ from \cite{Hamada:2019ack} in the way the energy-momentum tensor in ten dimensions is computed. This is also discussed in reference \cite{Carta:2019rhx}.

We demonstrate the difference in the language of \cite{Hamada:2019ack}. There the trace reversed Einstein equations are computed with variations w.r.t. the volume instead of the warp factor. Below we mention why this choice is important to reach the results presented in \cite{Hamada:2019ack}. Concretely, the 4d curvature can be computed by using the following equation \cite{Hamada:2019ack}\footnote{We use a different convention for the Einstein-Hilbert action than \cite{Hamada:2019ack}. This amounts for a difference of a factor 2 in the RHS of \eqref{eq:EinsteinEqHebecker}.}:
\begin{equation}
\label{eq:EinsteinEqHebecker}
	2\pi \mathcal{V}_6 \mathcal{V}_4 R_4 = -\left((\Im \rho) \frac{\delta}{\delta(\Im \rho)} - 1\right)S.
\end{equation}
We only look at the terms in the action involving the gaugino condensate where we integrated over 6D space:
\begin{equation}
\label{eq:ActionHebecker}
	S_{\lambda\lambda} = 2\pi\int \star_4 \frac{1}{2}\left[-8(\Im \rho)^{\frac{1}{2}} \left|\frac{\lambda\lambda}{16\pi^2}\right|^2+(\Im \tau)^{-\frac{1}{2}}\left((\Im \rho)^{\frac{1}{4}}\frac{\bar{\lambda}\bar{\lambda}}{16\pi^2}W_0+ c.c.\right)\right]
\end{equation}
Here we have written the volume dependence coming from the 6D metric explicitly and ignored the warping dependence.

As we have mentioned, our approach is as follows: we vary the D7 brane action to obtain its energy-momentum tensor. \emph{Subsequently} we fill in the vev of the gaugino condensate $\langle \lambda\lambda \rangle$, whose value is set by non-perturbative physics. This is how the semi-classical limit is usually defined: one inserts the quantum vevs of operators into the classical EM tensor. Following this approach, one finds by evaluating \eqref{eq:EinsteinEqHebecker} and \eqref{eq:EinsteinEqHebecker} that the 4d curvature satisfies:
\begin{equation}
	\mathcal{V}_6 R_4 = -2(\Im \rho)^\frac{1}{2}\left|\frac{\lambda\lambda}{16\pi^2}\right|^2 + \frac{3}{4}(\Im \tau)^{-\frac{1}{2}}\;\Re\left((\Im \rho)^{\frac{1}{4}}\frac{\bar{\lambda}\bar{\lambda}}{16\pi^2}W_0\right)
\end{equation} 
This result agrees with \eqref{GKP_regularised}.

However a subtlety arises which relates to the dependence of the gaugino vev on the gauge coupling  (\ref{vev}). Since the  coupling equals the inverse volume of the 4-cycle wrapped by the D7 branes, a dependence of the vev $\langle \lambda\lambda \rangle$ on the ten-dimensional metric is induced. This dependence cannot be understood classically in four or ten dimensions. One could imagine reversing the order and plug the metric dependence of the gaugino condensate into the ten-dimensional action \emph{before} computing the energy-momentum tensor. This introduces extra terms due to derivatives on the condensate, i.e. this approach would result in a 4d curvature of the form.
\begin{align}\label{hamadafineq}
\begin{split}
		\mathcal{V}_6 R_4 =&-2(\Im \rho)^\frac{1}{2}\left|\frac{\lambda\lambda}{16\pi^2}\right|^2 + \frac{3}{4}(\Im \tau)^{-\frac{1}{2}}\;\Re\left((\Im \rho)^{\frac{1}{4}}\frac{\bar{\lambda}\bar{\lambda}}{16\pi^2}W_0\right)\\		
		&+4(\Im \rho)^{\frac{3}{2}} \frac{\partial}{\partial\Im \rho}\left|\frac{\lambda\lambda}{16\pi^2}\right|^2-(\Im \tau)^{-\frac{1}{2}}\; \Re \left((\Im \rho)^{\frac{5}{4}}\left(\frac{\partial}{\partial \Im \rho}\frac{\bar{\lambda}\bar{\lambda}}{16\pi^2}\right)W_0\right)
\end{split}
\end{align}
Working with this equation, reference \cite{Hamada:2019ack} finds a qualitative match with the four-dimensional KKLT potential and further concludes there is no obstruction to uplifting to meta-stable dS vacua. We want to emphasize that that it is not clear how the semi-classical method really works in this case. %In particular, since replacing $\langle \lambda\lambda\rangle$ with the gaugino vev effectively eliminates 10d covariance, the method of \cite{Hamada:2019ack} is sensitive how exactly the variation of the action is performed so that the result is \eqref{hamadafineq}.
%A variation with respect to a different metric mode, such as the warp factor, presumably does not yield the same result. 
%A straightforward method that manifestly respects ten-dimensional covariance is to have no metric dependence in $\langle \lambda\lambda \rangle$ at the level of the 10d action.

In a rather different wording reference \cite{Carta:2019rhx} warned for this sublety and concludes that perhaps the ten-dimensional approach is not attainable after all. We remain agnostic about that point but emphasize the following; \emph{if} a ten-dimensional viewpoint is to make sense then we want to emphasize the following two things:
%that our approach is the one we have followed for the following reasons
 1) Inserting the four-dimensional information about the gauge coupling dependence into the ten-dimensional action implies one is not following a genuine ten-dimensional and local approach after all. 2) A semi-classical limit should be something universal. In other words when one couples classical gravity to quantum matter then the approach is that the quantum vevs are inserted into the classical EM tensor. The latter tensor is a universal object in the theory and how to compute the tensor does not depend on the quantum effects of the matter sector.

This does not imply we disagree with the outcome of the results in \cite{Hamada:2019ack}, but we want to emphasize our motivations for following the other path and what the consequences are.

Finally note that, if the second part of reference \cite{Carta:2019rhx} turns out to be correct, then dS uplifts are questionable already for a simpler reason related to having controlled uplifts and throat volumes that fit into the compact CY space.

\section*{Acknowledgments}
We acknowledge useful discussions with Arthur Hebecker, Jakob Moritz, Pablo Soler, Bert Vercnocke and Marco Scalisi. We are especially grateful for the help of Toine Van Proeyen in double checking various parts of the computations done in this paper.  TVR would like to thank the Center For The Fundamental Laws Of Nature at Harvard University for hospitality while this work was being finished. FFG is a Postdoctoral Fellow of the Research Foundation-Flanders. The work of VVH is supported by the by the European Research Council grant no.~ERC-2013-CoG 616732 HoloQosmos and the C16/16/005 grant of the KULeuven.  The  work  of  VV  is  supported  by  the  European  Research  Council grant  ERC-2013-CoG  616732  HoloQosmos.  The work of TVR is supported by the FWO odysseus grant G.0.E52.14 and the C16/16/005 grant of the KULeuven. 

\appendix

\section{Conventions}
The type IIB supergravity action (in units where $2\pi \ell_s =1 $) is
\be
S = 2\pi\int\star_{10}\left(R_{10} - \f{\dd\tau \cdot\dd \bar\tau}{2(\Im\tau)^2} - \f{G_3\cdot \overline{G_3}}{2(\Im\tau)} - \f14 |F_5|^2\right) - \f{\pi}{2i}\int \f{C_4\w G_3\w \overline{G_3}}{\Im\tau}~,
\ee
where $R_{10}$ is the ten-dimensional Ricci scalar calculated using the metric $g_{MN}$ with determinant $g_{10}$. The axion-dilaton is denoted by $\tau = C_0+i\e^{-\phi}$ and the NSNS and RR 3-form field strengths have been combined into a single complex 3-form 
\be
G_3 = F_3 - i\e^{-\phi} H = \dd C_2 - \tau \dd B_2~,\quad H=\dd B_2~,\quad F_3 = \dd C_2 - H C_0~.
\ee
Throughout the paper we use short-hand notation to denote form contractions, let $\omega_p$ and $\psi_p$ denote two $p$-forms, then
\be
\star_{10}\omega_p \w \psi_p = \omega_p\cdot \psi_p~\star_{10}1~,\quad \omega_p\cdot \psi_p = \f{1}{p!}\omega_{M_1M_2\cdots M_p}\psi^{M_1M_2\cdots M_p}~,\quad |\omega_p|^2 =\overline{\omega}_p\cdot\omega_p~.
\ee
The complex-conjugation in the last expression is to allow for the possibility that $\omega_p$ is a complex $p$-form.

%\section{The projector $\mathcal{P}$}
%\label{projectorsec}
%Any form $\omega$ can be decomposed into three orthogonal terms by hodge decomposition: an exact term term $d\alpha$, an harmonic term $\gamma$, and a co-exact term $d^\dagger \beta$,
%\be
%\omega = d \alpha + \gamma + d^\dagger \beta \,.
%\ee
%We define the projector $\mathcal{P}$ to act by removing the co-exact term,
%\be
%\mathcal{P}(\omega) \equiv d \alpha + \gamma \,.
%\ee
%Note that $\mathcal{P}$ maps forms to the kernel of the de Rham differential.
%
%\VVH{The following text does not belong here, but we maybe should explain how it works with our $I$.}
%If a form has a definite scaling with respect to a variable, then every term it is composed of must scale the same way with respect to that variable. If some term had a different scaling, the form as a whole would not have a definite scaling. So, if $\omega$ has a definite scaling with respect to the warp factor, it follows that all its terms under hodge decomposition obey the same scaling with respect to the warp factor. It is thus transparent that $\mathcal{P}$ has no effect on the scaling of a form with respect to the warp factor.

\section{Variation of the action}
\label{Appendix_warp_variation}
In this appendix we derive eq. \eqref{Corrected_GKP_equation} step-by-step. We also fill in some of the gaps of the subsequent calculations in \cref{4dCC}. First we show that we can obtain equation \eqref{GKP} directly by varying the ten-dimensional type IIB supergravity action with respect to the warp factor $A$, as defined in \eqref{metric}. After establishing this, we include the D7-branes and redo the same computation.

The ten-dimensional action for the Ansatz (\ref{metric}, \ref{Ansatz2}) can be found to be:
\be
S_{\text{IIB}} = 2\pi \int \dd^{10}x \sqrt{-g_{4}}\sqrt{g_6} e^{-2A} \,\mathcal{I}\,,
\ee
where 
\be
\mathcal{I}=e^{-2A}R_4 + e^{2A}\left(R_6 + 2\triangle_6 A -8|\dd A|_6^2) \right)
-e^{2A}\frac{|\dd \tau|_6^2}{2(\text{Im} \tau)^2} -e^{6A}\frac{|{G_3}|_6^2}{2\text{Im} \tau} +e^{-6A}\frac{|{\dd}\alpha|_6^2}{2}\,.
\ee
The $|\cdots|_6^2$ denote metric contractions using purely the six-dimensional metric $\dd s_6^2$ without the warp factor.
We now compute a variation with respect to $A$. We find:
\be
\label{AvariationIIB}
0= R_4 - \triangle_6 e^{4A} + e^{-6A}|\dd e^{4A}|^2 + e^{2A}\frac{|G_3|^2}{2\text{Im} \tau} + e^{-6A}|\dd\alpha|^2\,.
\ee
This is simply the trace-reversed ten-dimensional Einstein-equation traced over the four-dimensional indices. Adding the Bianchi identity, we recover equation \eqref{GKP}:
\be
\label{GKP_equation_Appendix}
\triangle_6 \Phi^- = R_4 +  e^{2A}\frac{|G_3^-|^2}{\text{Im} \tau} + e^{-6A}|\dd \Phi^-|^2\,.
\ee
We did not include any sources, but including O7/O3 planes does not modify the final result. Now we would now like to add D7-branes to the configuration.
As explained in the main text, the inclusion of the D7-brane implies that we should perform the replacement \eqref{replacementG3toPerfectSquare} and vary that with respect to the warp factor. 
We have two crucial remarks. The careful reader will notice that $I$ as defined in \eqref{definition_I} contains a warp factor dependence of $e^{-4A}$. This factor should \textit{not} be varied when deriving the equivalent to \eqref{AvariationIIB}, since it does not arise from a metric in the action. On the other hand, the delta function and the holomorphic three-form do have a warp factor dependence. The holomorphic three-form goes like $e^{-3A}$ and the delta function like $e^{2A}$. We can therefore write the perfect square action as 
\begin{align}
S_{3} = -2\pi \int \dd^{10}x \sqrt{-g_{4}}\sqrt{g_{6}} \frac{e^{4A}}{2 \Im \tau} \left|{G_3} - e^{-A}(\Im \tau) \tilde{\mathcal{P}}(\bar{I})\right|_6^{2}\,,
\end{align}
where we have made the warp-factor dependence explicit by momentarily writing ${\cal P}(I) = e^{-A} \tilde{\mathcal{P}}(\bar{I})$. Using this we find
\begin{align}
\label{AvariationD7G3notrewritten}
-\frac{e^{4A}}{8\pi \sqrt{-{g}_{4}}\sqrt{\tilde{g}_{6}}} \frac{\delta S_3}{\delta A} &= \frac{e^{2A}}{2 \Im \tau}\left|G_3 - (\Im \tau)\mathcal{P}(\bar{I}) \right|^{2}+ \frac{e^{2A}}{4}\Re\left[\left(G_3 - (\Im \tau)\mathcal{P}(\bar{I})\right)\cdot\mathcal{P}(I)\right]\\
\label{AvariationD7G3}
&= e^{2A}\left(\frac{\left|G_3 \right|^{2}}{2 \Im \tau} - \frac{3}{4} \Re \left( G_3 \cdot \mathcal{P}(I)\right) + \frac{\Im \tau}{4} |\mathcal{P}(I)|^2\right)\,.
\end{align}
We should replace the $|G_3|^2$-term in \eqref{AvariationIIB} by \eqref{AvariationD7G3}. This modifies \eqref{GKP_equation_Appendix} to exactly \eqref{Corrected_GKP_equation} which reads:
\begin{equation}
\nonumber
\triangle_6 \Phi^- ={R}_4 + e^{-6A}|\dd \Phi^-|^2  + e^{2A}\left(\f{|G_3^-|^2}{ \Im \tau} - \f{3}{4}\Re \big(G_3\cdot {\cal P}(I)\big) + \f{\Im\tau}{4} |{\cal P}(I)|^2\right)~.
\end{equation}

If one now writes $\mathcal{P}(I)= I^{\text{sing}} + I^{\text{reg}}$ as in \eqref{defIsingIreg} and substitutes the solution to the equation of motion $G_3 = (\Im \tau)\overline{I^{\text{sing}}} + G_3^0$ (while making the approximation that $\Phi^{-}=0$ and warping, $C_4$-potential and axion-dilaton are constant) into \eqref{AvariationD7G3notrewritten}, one finds together with the other terms in the IIB action and the Bianchi identity the following equation:
\begin{align}
\begin{split}
0 =&\, R_4 + \frac{e^{2A}}{2 \Im \tau}\left|G_3^0 - (\Im \tau)\overline{I^{\text{reg}}}\right|^2+ \frac{e^{2A}}{4 \Im \tau}\Re\biggl[\left(G_3^0 - (\Im \tau)\overline{I^{\text{reg}}}\right) \cdot (\Im \tau)\left(I^{\text{sing}} + I^{\text{reg}}\right)\biggr]\\
& + \frac{e^{2A}}{2 \Im \tau} \biggl(|G_3^{0,-}|^2 + 2\Re\left(G_3^{0,-}\cdot\overline{I^{\text{sing},-}}\right) - |G_3^{0,+}|^2 - 2\Re\left(G_3^{0,+}\cdot \overline{I^{\text{sing},+}}\right)\biggr)
\end{split}
\end{align}
The substitution already took into account that the Bianchi identity does not contribute $\delta^2$ singularities, which is discussed in \cref{CSappendix}.
Recall that  $I^{\text{sing}}$ is by definition exact, so its inner product with a harmonic form ($G^0_3$ or $I^{\text{reg}}$) is therefore zero when the expression is integrated over (the (A)ISD-parts vanish in the same fashion).
%\footnote{\color{red} Something about the fact that we remove singularities in the same way as \cite{Hamada:2018qef} removes the co-exact piece.}
In this way, we arrive at
\be
0 =\int\star_6\left( {R}_4  + \frac{e^{2A}}{ \Im \tau}|G_3^{0,-}|^2 -\frac{3e^{2A}}{4}\Re\left(G_3^0  \cdot I^{\text{reg}}\right) + \frac{e^{2A}}{4}(\Im \tau) |I^{\text{reg}}|^2\right)~.
\ee
%{\color{red} {\color{blue} Should we include this here?}
%Evaluating the last term, we find that
%\begin{align}
%	\int \tilde{\star}_6\frac{e^{2A}}{4}(\Im \tau) |I^{\text{reg}}|^2 = \frac{1}{4}\left|\frac{\lambda\lambda}{16\pi^2} \right|^2 \int \frac{1}{\mathcal{V}_2^2} \tilde{\star}_6 \Omega \wedge \bar{\Omega}
%\end{align}
%With $-N/32\pi^2 \lambda \lambda = \mathcal{A}e^{ia\rho}$, we can write
%\begin{align}
%\int \tilde{\star}_6\frac{e^{2A}}{4}(\Im \tau) |I^{\text{reg}}|^2 = \frac{1}{N^2} \left|\mathcal{A}e^{ia\rho}\right| \int \frac{1}{\mathcal{V}_2^2} \tilde{\star}_6 \Omega \wedge \bar{\Omega}
%\end{align}
%The KKLT value for the potential at the AdS minimum is
%\begin{align}
%	V_{AdS} = -\frac{1}{6 L^{4}}a^2 \left|\mathcal{A}e^{ia\rho}\right|
%\end{align}
%}

\section{(Non)renormalisation of the Chern-Simons action}
\label{CSappendix}

The effective bulk action of IIB string theory contains a Chern-Simons term,
\be
S_{\text{CS}} = -\frac{\pi}{2i} \int \frac{C_4 \wedge G_3 \wedge \overline{G_3}}{\Im \tau} \,.
\ee
This term is topological and does not affect the Einstein equation directly. Still one can be concerned and wonder whether we should renormalise the Chern-Simons action or not. If we should, the renormalisation could introduce metric dependence and affect our results.

Under the assumptions of constant $C_4$ and axion-dilaton made in the main body of the text, we should not renormalise the Chern-Simons term. Not renormalising produces no disasters and thus there is perhaps no obvious need for renormalisation. Let us check this.

At first, it may appear that filling the solution for $G_3$ into the Chern-Simons action will produce a badly divergent action. However, this is not the case. Filling the on-shell value of $G_3$ into the Chern-Simons action yields
\begin{align}
S_{\text{CS,on-shell}} = &-\frac{\pi g_s}{2i} \int \Biggl\{C_4 \wedge G_3^0 \wedge \overline{G_3^0} + (\Im \tau) C_4 \wedge G_3^0 \wedge \overline{I^{\text{sing}}} + (\Im \tau)  C_4 \wedge I^{\text{sing}} \wedge \overline{G_3^0} \Biggr\} \nonumber\\
&-\frac{\pi g_s}{2} \int \left(|I^{\text{sing} \, -}|^2 - |I^{\text{sing} \, +}|^2  \right) (\Im \tau)^2 C_4 \wedge \tilde{\star}_6 1 \,, \label{CSonshellwithdiv}
\end{align}
where we introduced the symbols $I^{\text{sing} \, \pm} = \frac{1}{2} (\star_6 \pm i) I^{\text{sing}}$.
The last two terms in this expression seem divergent while all other terms are clearly finite upon evaluation of the integral. Interestingly the two divergent terms exactly cancel against each other and the on-shell action is well-behaved. To see this, note that near a D7-brane we can locally write the internal manifold as $\mathcal{M}=\Sigma_2 \times \Sigma_4$, with $\Sigma_2$ transverse and $\Sigma_4$ parallel to the D7-brane. We then have 
\begin{align}
\overline{I^{\text{sing}}} &\sim \dd\left(\frac{\partial G}{\partial z} \right) \wedge \bar{\Omega}_2\nonumber\\
&= \left(\frac{\partial^2 G}{\partial z\partial z} \right)\dd z \wedge \bar{\Omega}_2 + \left(\frac{\partial^2 G}{\partial z \partial \bar{z}} \right) \dd\bar{z} \wedge \bar{\Omega}_2
\label{Isingdecompose}
\end{align}
with $G$ the Green's function, which varies transverse to the D7-brane, and $\dd z$ and $\dd \bar{z}$ are defined on $\Sigma_2$. It is clear that the first of these terms is $(1,2)$ while the second is $(0,3)$. First we notice that the two terms have the same magnitude:
\be
\int\star_6 \left|\left(\frac{\partial^2 G}{\partial z\partial z} \right)\dd z \wedge \bar{\Omega}_2 \right|^2 = \int\star_6 \left|\left(\frac{\partial^2 G}{\partial z \partial \bar{z}} \right) \dd\bar{z} \wedge \bar{\Omega}_2 \right|^2 \,,
\ee
which follows from integration by parts and the reality of $G$. Then we use that the first term in \eqref{Isingdecompose} corresponds to the IASD part of $\overline{I^{\text{sing}}}$, while the second term corresponds to the ISD part. It is immediate that the $(0,3)$ term is ISD as the unique $(0,3)$ form at our disposal is $\bar{\Omega}$ and this is ISD in our convention. That the $(1,2)$ form is IASD follows from the fact that it is primitive and primitive $(1,2)$ forms are IASD in our convention. The primitivity can be seen by decomposing $J$ as
\be
J = J_{\Sigma_2} + J_{\Sigma_4} \,,
\ee
with the two terms defined in the obvious way.
Using this, evaluation yields
\be
\left(\frac{\partial^2 G}{\partial z\partial z} \right)\dd z \wedge \bar{\Omega}_2\wedge J = 0 \,.
\ee
We thus see that  $I^{\text{sing}\, +}$ and $I^{\text{sing}\, -}$ have the same magnitude. Therefore, the two problematic terms in \eqref{CSonshellwithdiv} cancel against each other and the on-shell Chern-Simons action is finite without the introduction of any renormalisation counterterms.

\bibliographystyle{utphys}

\bibliography{refs}

\end{document}